\documentclass{elsart}

\usepackage{amssymb}
\usepackage{amsmath}
\usepackage{amsfonts}

\renewcommand\ln{\ell{n}}
\newcommand\beq{\begin{equation}}
\newcommand\eeq{\end{equation}}
\newcommand\bea{\begin{eqnarray}}
\newcommand\eea{\end{eqnarray}}
\newcommand\bseq{\begin{subequations}} 
\newcommand\eseq{\end{subequations}}
\newcommand\bal{\begin{align}}  
\newcommand\ealign{\end{align}}    

\renewcommand\ln{{\rm ln}}

\begin{document}

\begin{frontmatter}



\title{Bianchi IX Chaoticity:  
BKL Map and Continuous Flow}

\author[FII,ICRA]{Giovanni Imponente%
	 \corauthref{cor1}}\ead{imponente@icra.it}
%
%
\author[ICRA,LASAP]{Giovanni Montani}\ead{montani@icra.it}
\address[FII]{Dipartimento di Fisica, 
 						Universit\'a ``Federico II'', and INFN Napoli -- Italy}
\address[ICRA]{ICRA -- International Center for
							Relativistic Astrophysics}
\address[LASAP]{Dipartimento di Fisica -- G9
					Universit\'a di Roma ``La Sapienza'', Roma -- Italy}

\corauth[cor1]{c/o G9 - Dip. 
Fisica, Universit\'a ``La Sapienza''
P.za A.Moro 5, I-00185 Roma}

\begin{abstract}
We analyze the Bianchi IX dynamics (Mixmaster) in view of 
its stochastic properties; in the present paper
we address either 
the original approach due to Belinski, Khalatnikov 
and Lifshitz (BKL) as well as a Hamiltonian one 
relying on the Arnowitt--Deser--Misner (ADM) reduction.\\
We compare these two frameworks and show how the BKL
map is related to the geodesic flow  associated with the 
ADM dynamics. 
In particular, the link existing  between the 
\textit{anisotropy parameters} and the 
\textit{Kasner indices} is outlined. 

\end{abstract}

\begin{keyword}
variational principles \sep ensemble theory \sep chaos \sep
early Universe

\PACS 
04.20.Fy \sep
05.20.Gg \sep
05.45.Gg \sep
98.80.Cq 

\end{keyword}
\end{frontmatter}

The Bianchi classification characterizes all the 
admissible three-dimensional homogeneous spaces and 
determines 9 non-equivalent types. 
The Bianchi mo\-dels were widely studied in cosmology 
as the simplest generalization of the 
Friedmann--Lemaitre--Robertson--Walker Universe, 
the Einstein equations reducing to an ordinary 
differential system with respect to time \cite{BKL70}. \\
Among the Bianchi classification, the types VIII 
and IX (Mixmaster) were particularly attractive 
from a cosmological point of view, being the 
most general ones allowed within the homogeneity 
constraint and characterized by a chaotic dynamics 
near the Big Bang; 
here we discuss the type IX in some detail, yet all 
considerations hold also for the type VIII.

The Bianchi IX model is summarized by the line element
\beq
ds^2= N^2(t) dt^2-\gamma_{ij}\sigma^i \sigma^j
\, , \qquad i,j=1,2,3
\label{lineel}
\eeq
where $N$ denotes the lapse function and the 
$\sigma^i$ are the 1-forms defined by the 
homogeneity property. 
In vacuum we can take
\beq
\gamma_{ij}(t) =\textrm{diag}
\left(e^{h_i(t)} \right)
\label{gammaij}
\eeq
and the Einstein equations provide
\bseq
\bal
& h^{\prime \prime}_i = \left( e^{h_j}-e^{h_k}\right)^2
	-e^{2 h_i} \, , \qquad i\neq j \neq k \label{ee1}\\
& \sum_i h^{\prime \prime}_i = 
	\frac{1}{2} \sum_{j \neq k} h^{\prime}_j h^{\prime}_k 
	\label{ee2}
\end{align}
\label{eee}
\eseq
where $(~)^{\prime}=d/d\eta$ and the time $\eta$ 
corresponds to the choice 
$N=\textrm{exp} 
\left(\sum_i \frac{h_i}{2}\right)$; 
equation (\ref{ee2}) plays the role of a first
integral for the system (\ref{ee1}).\\
Belinski, Khalatnikov and Lifshitz (BKL) \cite{BKL70}
showed 
that the asymptotic approach to the Big Bang of the 
type IX model is characterized by an infinite sequence
of \textit{Kasner epochs}, i.e. intervals of time 
during which the right-hand side of (\ref{ee1}) is 
negligible. Thus, in the limit 
$\eta \rightarrow - \infty$, we have a piecewise 
solution, whose single steps take the Kasner form
\beq
h_i \sim p_i \eta + h^{*}_i \, , 
\qquad \quad (p_i, h^{*}_i)=\textrm{const.} \, , 
\label{ksol}
\eeq
where the Kasner indices $p_i$ satisfy the conditions
$\sum_i p_i = \sum_i (p_i)^2 =1$. \\
The Mixmaster dynamics is then reduced to a map 
on these indices: 
adopting the parametrization which orders 
the $p_i$'s as
\beq
p_1 = -\frac{u}{1+u+u^2}\, , \quad
p_2 = \frac{1+u}{1+u+u^2}\, ,\quad
p_3 = \frac{u(1+u)}{1+u+u^2}\, ,\,
\quad u\in (1,\infty)
\label{ki}
\eeq
then we get the following discrete map on the 
parameter $u$
\beq
u^{\prime} = \left\{ 
				\begin{array}{l}
				u-1 					\qquad 	 u>2 \\
				\displaystyle \frac{1}{u-1} \qquad    1<u \leq 2
				\end{array}
					\right.
\label{umap}
\eeq
The main achievement of the BKL approach is the 
discovery that such a map has stochastic properties
and $u$ admits the probability distribution $w$
\beq
w(u) = \frac{1}{u(u+1) \ln 2} \, .
\label{uprdist}
\eeq
In \cite{BKL82} the above analysis was extended, 
point by point in space, to a generic inhomogeneous 
cosmological model.\\
To get a continuous description, let us introduce
the Misner variables $(\alpha,\beta_+,\beta_-)$
and then the Misner--Chitre-like ones $(\tau, \xi, \theta)$, 
via the transformations
\bseq
\bal
& h_1= \beta_+ + \sqrt{3} \beta_-\, , \quad
  h_2= \beta_+  - \sqrt{3} \beta_-\, ,  \quad
  h_3= -2 \beta_+ \\
& \alpha =- e^{\tau}\xi\, , \qquad
  \beta_{\pm} = e^{\tau}\sqrt{\xi^2 -1} \times
  \left\{ \begin{array}{l}
  				\cos \theta \\
  				\sin \theta
  				\end{array} \right.
\end{align}
\eseq
with $\xi \in (1, \infty)$ and
$\theta \in [0, 2\pi)$. \\
By an Arnowitt--Deser--Misner (ADM) reduction 
of the Einstein--Hilbert action, it can be shown \cite{IM01}
that the Bianchi IX dynamics is summarized by 
\beq
S_{\textrm{IX}}= \int d\tau \left(
  p_{\xi} \frac{d\xi}{d\tau} +
  p_{\theta} \frac{d\theta}{d\tau} -
  \sqrt{\varepsilon^2 + U} \right) \, ,
\label{b9action}
\eeq
where $\varepsilon^2 = \left( \xi^2 -1 \right) p^2_{\xi}+
\frac{p^2_{\theta}}{\xi^2 -1 }$.\\
Near the Big Bang $(\tau \rightarrow  \infty)$, 
the potential is modelled by the infinite walls
\beq
U= \sum_i \Theta_{\infty}(H_i) \, , \quad 
	\Theta_{\infty}(x)= \left\{ \begin{array}{l}
												0 			\quad x>0 \\
												\infty  \quad x\leq 0
												\end{array}
												\right.
\label{pinf}
\eeq
where the quantities 
$H_i \equiv h_i/(\sum_k h_k)$, 
$(\sum_i H_i =1)$ are the  
\textit{anisotropy para\-me\-ters} which in the 
Misner--Chitre-like variables read as
\bseq
\bal
& H_1 = \frac{1}{3} - \frac{\sqrt{\xi ^2 - 1}}{3
		\xi }\left(\cos\theta + \sqrt{3}\sin\theta \right)  \\ 
& H_2 = \frac{1}{3} - \frac{\sqrt{\xi ^2 - 1}}{3
		\xi }\left(\cos\theta - \sqrt{3}\sin\theta \right)  \\ 
& H_3 =  \frac{1}{3} + 2\frac{\sqrt{\xi ^2 - 1}}{3\xi }
       \cos\theta \, ;
\end{align}
\label{hs}
\eseq
the functions $H_i$ expressed in (\ref{hs}) 
do not depend on the (time) variable $\tau$. \\
The domain $\Gamma_H$ where $U$ vanishes is dynamically
closed and, within it, $\varepsilon$ behaves as a constant
of motion, i.e. 
$	\frac{d\varepsilon}{d\tau}=
\frac{\partial\varepsilon}{\partial\tau}=0$, which 
implies $\varepsilon=E=\textrm{const.}$. \\
Thus the Bianchi IX dynamics is isomorphic to a 
billiard on a Lobachevski plane whose line element
reduces to 
\beq
dl^2 = E^2 \left[ \frac{d\xi^2}{\xi^2 -1} +
    (\xi^2 - 1)d \theta^2 \right] \, . 
\label{lineb}
\eeq
This system is stochastic \cite{H94,IM01}
and, upon the transformation 
of the conjugate momenta as
\beq
p_{\xi} =\varepsilon \frac{\cos \phi}{\sqrt{\xi^2-1}} \, ,
 \quad
p_{\theta} =\varepsilon \sqrt{\xi^2-1}~{\sin \phi} \, ,
\qquad 0\leq \phi < 2\pi \, ,
\eeq
the dynamics admits a uniform 
\textit{invariant measure} like \cite{CB83,KM97}
\beq
d\mu = \frac{1}{8\pi^2} d\xi d\theta d\phi \, .
\eeq
The link between this approach based on continuous 
variables and the one based on the discrete BKL one
can be achieved considering the following set of 
coordinates transformations
\bseq
\bal
\vec{y} &= \frac{1+\xi}{\sqrt{\xi^2 -1}} 
(\cos \theta, \sin \theta) \, ,  \\
\vec{\iota} &= \frac{\vec{y}+\vec{b}}{ 
	(\vec{y}+\vec{b})^2} - \vec{b} \, ,
\end{align}
\eseq
where $\vec{\iota}$ denotes the Poincar\'e 
variables and $\vec{b}$ belongs to the 
boundary of the Lobachevsky plane, 
i.e. $b^2=1$; let us consider afterwards
the transformation to $(u,v)$
\bal
\vec{\iota} = \frac{2}{\sqrt{3}} &\left[ \left( 
		u + \frac{1}{2} \right) \vec{b}^{\bot} +
		v \vec{b} \right] \, , \\
		& v\geq 0 \, , \quad
		- \infty < u < \infty \, , \quad 
		\vec{b}=(0,1) \,, \quad \vec{b}^{\bot}=(1,0) \, .\nonumber
\end{align}
In terms of the $(u,v)$ variables, the anisotropy 
parameters (\ref{hs}) read
\bal
Q_1(u,v)= -\frac{u}{\delta} \, , \quad
Q_2(u,v)= \frac{1+u}{\delta} \, , \quad
Q_3(u,v)= \frac{u(u+1)+ v^2}{\delta} \, ,
\label{quv}
\end{align}
with $\delta(u,v)\equiv (u^2 + u +1 + v^2)$. 
The quantitities in (\ref{quv}) reduce to 
the Kasner exponents (\ref{ki}) when evaluated 
at $v=0$. \\
This correspondence provides the link between 
the dynamics representation in continuous variables 
and the BKL map \cite{KM97}.



\end{document}